\documentclass[12pt]{article}
\usepackage{epsf}
\usepackage{cite}
\usepackage{amsmath}
\usepackage{graphics}
\usepackage{color}
\input{colordvi.tex}

\renewcommand{\thefootnote}{\#\arabic{footnote}}

\newcommand{\gtrsim}{\mathop{}_{\textstyle \sim}^{\textstyle >}}
\newcommand{\lesssim}{\mathop{}_{\textstyle \sim}^{\textstyle <} }

\begin{document}

\setcounter{footnote}{0}

\renewcommand{\thepage}{\arabic{page}}
\setcounter{page}{1}
\renewcommand{\thefootnote}{\#\arabic{footnote}}

\begin{titlepage}

\begin{center}

\hfill TU-851 \\
\hfill September, 2009\\

\vskip .5in

{\Large \bf
Cosmology with Long-Lived Charged \\ Massive Particles
}

\vskip .45in

{\large
Kazunori Kohri$^{1,2}$ and Tomo Takahashi$^3$
}

\vskip .45in

{\em
$^1$
Department of Physics, Tohoku University, Sendai 980-8578, Japan\\
\vspace{2 mm}
$^2$
Physics Department, Lancaster University, Lancaster LA1 4YB, UK\\
\vspace{2 mm}
$^3$
Department of Physics, Saga University, Saga 840-8502, Japan \\
}

\end{center}

\vskip .45in

\begin{abstract}
  We investigate the evolution of the bound state of negatively
  charged massive particles (CHAMPs) with light elements and discuss
  its cosmological consequences and the constraint.  By numerically
  solving the Boltzmann equation, we study the time evolutions of such
  bound states.  Since most of negative CHAMPs are captured by
  $^4\mbox{He}$, its bound state is positively charged and couples
  with the electromagnetic plasma.  When charged particles constitute
  a dominant non-relativistic component, density fluctuations of
  matter cannot grow due to the acoustic damping.  This results in the
  suppression of matter power spectrum from which a severe constraint
  can be obtained.  By arguing constraints from other aspects of
  cosmology, we show that the constraint from large scale structure
  gives most stringent one in some representative cases.
\end{abstract}

\end{titlepage}

\section{Introduction}

Long-lived charged massive particles (CHAMPs) can exist in 
various extensions of the standard model of particle physics such as
supersymmetry (SUSY). One of such example is 
a slepton, a superpartner of leptons, which can be 
stable if it is the lightest supersymmetric particle (LSP)  and R-parity is conserved.
However, the abundance of such stable charged massive particles 
would be severely constrained \cite{Kudo:2001ie}, in particular, 
from experiments of the deep sea water
\cite{Smith:1979rz,Smith:1982qu,Hemmick:1989ns,Verkerk:1991jf,
  Yamagata:1993jq,Amsler:2008zzb}. Although there might
have been a mechanism to prevent them being captured into the
Earth~\cite{Chuzhoy:2008zy} and in such a case the constraints may not
be applicable, a scenario with stable CHAMPs would generally not be
viable.  However, in some other scenarios, CHAMPs are unstable, and
they can constitute a dominant component of non-relativistic particles
in the early Universe.  For example, when the gravitino is the LSP,
which can be easily realized in gauge-mediated SUSY breaking models
\cite{Dine:1994vc,Dine:1995ag,Giudice:1998bp}, the next lightest
supersymmetric particles (NLSP) may be CHAMPs, and they can be
long-lived. Although such unstable CHAMPs can evade the constraint
from the sea water, they affect other aspects of cosmology.  It has
been rigorously investigated that the decay of such massive particles
would destroy light elements synthesized by big bang nucleosynthesis
(BBN), from which we obtain constraints on the properties of CHAMPs
such as the decay rate and its abundance
\cite{RadDecOld,Khlopov:1999rs,Dimopoulos:1987fz,Kawasaki:1994af,
  Jedamzik:1999di,Kawasaki:2000qr,Kohri:2001jx,Cyburt:2002uv,Kawasaki:2004yh,
  Ellis:2005ii,Jedamzik:2006xz,Cyburt:2009pg}.  Although the
considerations of the decay also applies to neutral massive particles,
there is another important effect on BBN which is specific to the
charged particles: the bound-state effect. Negative CHAMPs can form a
bound state with positively charged light elements, which affects the
BBN reaction rates and their abundances \cite{CHAMPold,Fargion:2005ep,
  Pospelov:2006sc,Kohri:2006cn,Kaplinghat:2006qr,Cyburt:2006uv,
  Steffen:2006wx,Hamaguchi:2007mp,Kawasaki:2007xb,Bird:2007ge,Pradler-Steffen,
  jittoh,jedamzikCBBN,Kusakabe:2007,
  Pospelov:2007js,Bailly:2008yy,CBBNsolution}.  In fact, as will be
shown later, most negative CHAMPs (referred to as $X^{-}$) are
captured by $^4$He, a double-positively charged element.  If this
bound state $(^4 {\rm He} X^{-})$ is stable for some time in the
course of the history of the universe, it would also affect other
aspects of cosmology in addition to BBN such as large scale structure.
Thus a detailed investigation of how the bound states are formed and
evolve would be important and interesting, which is one of the main
topics in this paper.
 
When a particle possesses an electric charge before recombination, it
is tightly coupled with the plasma of electrons and photons. Thus such
charged particles, which are supposed to be the bound state $(^4 {\rm
  He} X^{-})$ here, can also participate in the acoustic oscillations.
When the bound state of negative CHAMPs constitutes a dominant
component of non-relativistic particles, fluctuations of
(non-relativistic) matter cannot grow due to the ``acoustic damping"
caused by the acoustic oscillations, which is different from the case
with a standard neutral cold-dark matter (CDM) model. This results in
the suppression of the matter power spectrum at some scales
\cite{Sigurdson:2003vy,Profumo:2004qt,Hisano:2006cj}.  Thus the bound
state $(^4 {\rm He} X^{-})$ should have a significant effect on large
scale structure, and the consideration of this issue can place a
unique bound on the properties of (negative) CHAMPs, which is another
topic we are going to focus in this paper.

The organization of this paper is as follows. In the next section, we
first carefully investigate the evolution of  the bound states of
$X^-$ with some light elements by numerically solving the Boltzmann
equation.  Then we discuss the effect of the (charged) bound state on
large scale structure and the damping of matter power spectrum.  In
Section~\ref{sec:const}, we discuss the constraint on the property of
CHAMPs from large scale structure and some other aspects such as BBN
and  the CMB spectrum. The final section is devoted to the summary of
this paper.

Unless otherwise stated, throughout this paper $n_{i}$, $m_{i}$ and
$Y_{i}$ denote the number density, the mass and the yield variable
($\equiv n_{i}/s$ with $s$ the entropy density) of a particle $''i''$,
respectively.

\section{Evolution of the bound state}\label{sec:bound}

We are interested in the bound-state formation of $X^{-}$ with a light
element, which occurs after the cosmic temperature becomes lower than
30 keV. Until that time, most of the standard BBN processes should
have almost been finished. Under this circumstance, the Boltzmann
equations for the time-evolution of the number density of bound states
$(N_{i}X^{-})$, denoted as $n_{(N_{i}X^{-})}$, with $N_{i} = p$ and
$^{4}$He are expressed by
\begin{eqnarray}
\label{eq:bolzhe4}
  \frac{d n_{(^4{\rm He}X^-)}}{dt} 
& = & - 3Hn_{(^4{\rm He}X^-)}
  -\Gamma_{X}n_{(^4{\rm He}X^-)} \nonumber  \\
  && +  \langle \sigma_{{\rm bnd},^{4}{\rm He}} v \rangle 
  \left[ (n_{\rm ^4He} - n_{(^4{\rm He}X^-)}) n_{X^{-}} 
   - \left(\frac{m_{\rm ^4He} m_{X} T}{2\pi m_{(^4{\rm He}X^-)}}\right)^{3/2}
   e^{-E_{b{^{4}{\rm He}}}/T}n_{(^4{\rm He}X^-)} \right] \nonumber \\
  && +  \langle \sigma_{{\rm ex}} v \rangle 
   (n_{\rm ^4He} - n_{(^4{\rm He}X^-)}) n_{(pX^-)}, \\
\label{eq:bolzP}
  \frac{d n_{(pX^-)}}{dt} & = &  - 3Hn_{(pX^-)}
  -\Gamma_{X}n_{(pX^-)} \nonumber\\
   && +  \langle \sigma_{{\rm bnd},p} v \rangle 
  \left[ (n_p - n_{(pX^-)}) n_{X^{-}} 
   - \left(\frac{m_p m_{X} T}{2\pi m_{(pX^-)}}\right)^{3/2}
   e^{-E_{bp}/T}n_{(pX^-)} \right]\nonumber \\
  && -  \langle \sigma_{{\rm ex}} v \rangle 
   (n_{\rm ^4He} - n_{(^4{\rm He}X^-)}) n_{(pX^-)},
\end{eqnarray}
where $n_{X^-}$ is the number density of free $X^{-}$, $\Gamma_{X}$ is
the decay width of $X$, $E_{b^{4}{\rm He}}\simeq 337.33\ {\rm keV}$
and $E_{bp}\simeq 24.97\ {\rm keV}$ are the binding energies of $({\rm
  ^4He}X^-)$ and $(pX^-)$ \cite{Hamaguchi:2007mp}, and $n_{\rm ^4He}$
and $n_p$ are the number densities of $^4$He and proton including both
free and bound-states.  The masses of the bound states are given by
$m_{({\rm ^4He}X^-)} = m_{\rm ^4He}+m_{X}-E_{b^{4}{\rm He}}$ and
$m_{(pX^-)} = m_p + m_{X}-E_{bp}$, respectively. The
thermally-averaged recombination cross sections $\langle \sigma_{{\rm
    bnd},^{4}{\rm He}} v \rangle$ and $\langle \sigma_{{\rm bnd},p} v
\rangle$ for the $^4$He and $p$ bound-states formation are given by
\cite{Kohri:2006cn}
\begin{eqnarray}
   \langle \sigma_{{\rm bnd},^{4}{\rm He}} v \rangle & \simeq &
   98.46
   \frac{\alpha E_{b^{4}{\rm He}}}{m_{\rm ^4He}^2\sqrt{m_{\rm ^4He} T}}, \\
   \langle \sigma_{{\rm bnd},p}  v \rangle & \simeq &
   24.62
   \frac{\alpha E_{bp}}{m_p^2\sqrt{m_pT}},
\end{eqnarray}
where $T$ is the cosmic temperature, and $\alpha$ is the fine
structure constant.

The terms in the third lines of the right-hand side of
Eqs.~(\ref{eq:bolzhe4}) and (\ref{eq:bolzP}) represent a
charge-exchange reaction,
\begin{eqnarray}
    \label{eq:CEX}
    (pX^{-}) + ^{4}{\rm He} \to (^{4}{\rm He}X^{-}) + p,
\end{eqnarray}
with its thermally-averaged cross section being denoted as $\langle
\sigma_{{\rm ex}} v\rangle$. This process is effective just after
$(pX^{-})$ has been formed.  Recently it has been reported that a rate
of this charge-exchange reaction is more rapid than the Hubble
expansion rate~\cite{Kamimura:2008fx} at the formation epoch of
$(pX^{-})$ which is approximately given by
\begin{eqnarray}
    \label{eq:rateratio}
    \frac{\langle \sigma v \rangle n_{^{4}{\rm He, free}}} {H}  \sim 2.6
    \times
10^{4}
    \left( \frac{T}{0.5 {\rm keV}} \right),
\end{eqnarray}
where we took the yield variable of free $^{4}$He to be $Y_{^{4}{\rm
    He,free}} = 5.4 \times 10^{-12}$. This means that the produced
$(pX^{-})$ is immediately destroyed, and rather $X^{-}$ is included
into $(^{4}{\rm He}X^{-})$ after the destruction.

\begin{figure}[t]
\begin{center}
\resizebox{70mm}{!}{\includegraphics{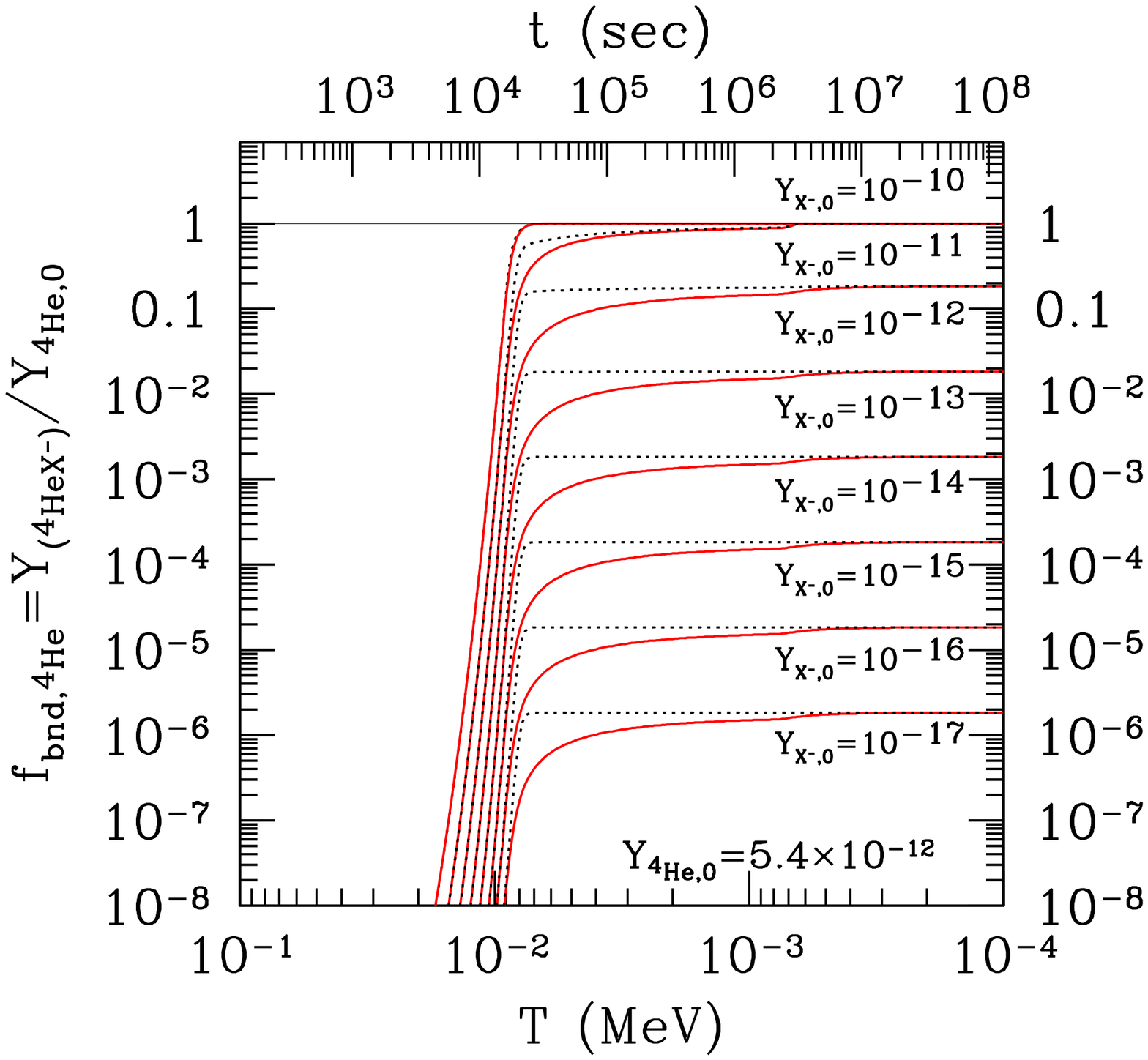}}
\resizebox{70mm}{!}{\includegraphics{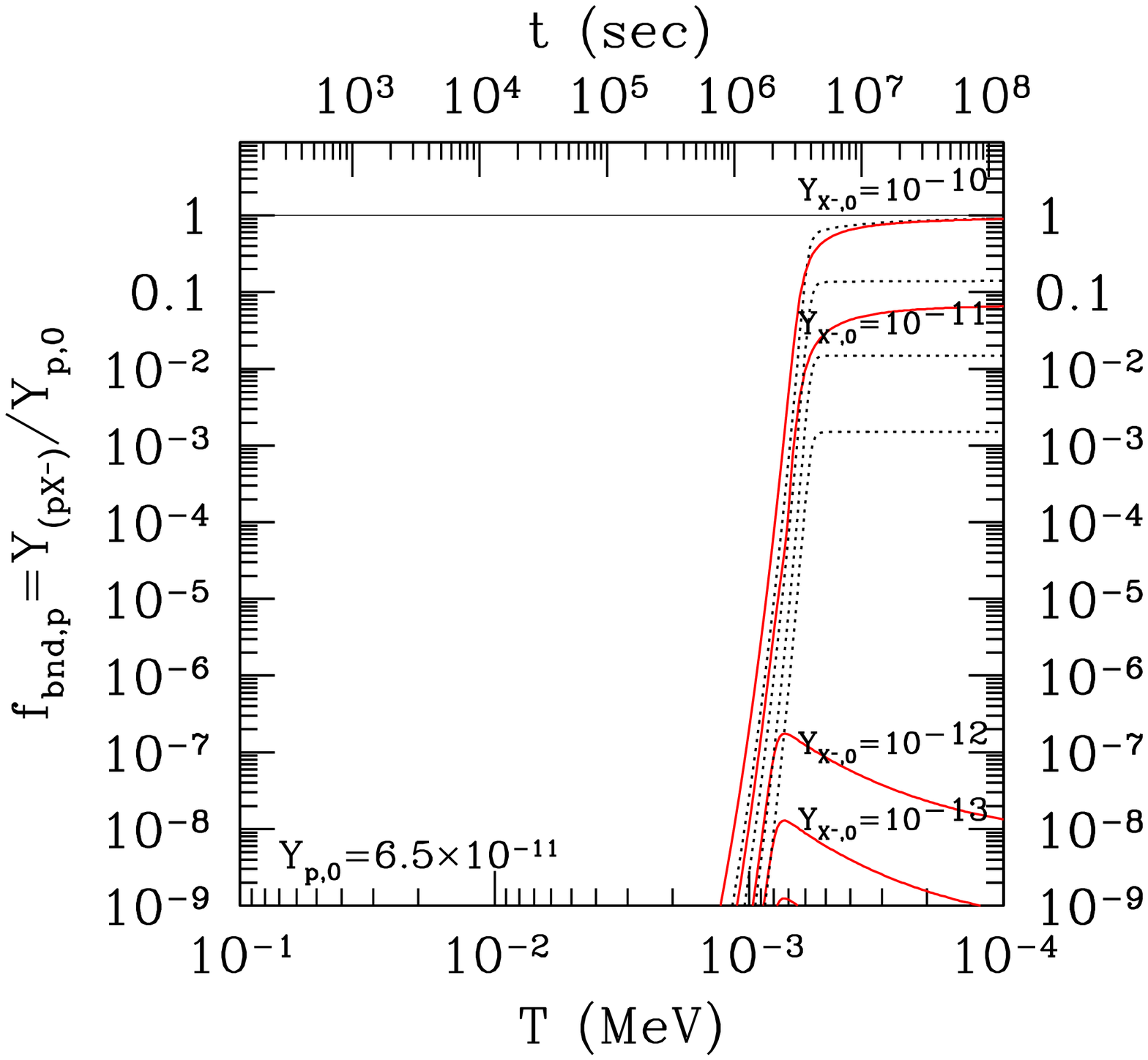}}
\caption{ Time evolution of $f_{\rm bnd, N_i} \equiv
  Y_{(N_iX^-)}/Y_{N_i,0}$ with $N_i=^4$He (left) and $N_i=p$ (right),
  where $Y_{N_i,0}$ and $Y_{X^{-},0}$ are the initial values of
  $Y_{N_i}$ and $Y_{X^{-}}$, respectively. The dashed lines denote
  solutions of Saha's equation. In these figures we assumed that
  $X^{-}$ is stable.}
\label{fig:fraction1}
\end{center}
\end{figure}

\begin{figure}[t]
\begin{center}
\resizebox{70mm}{!}{\includegraphics{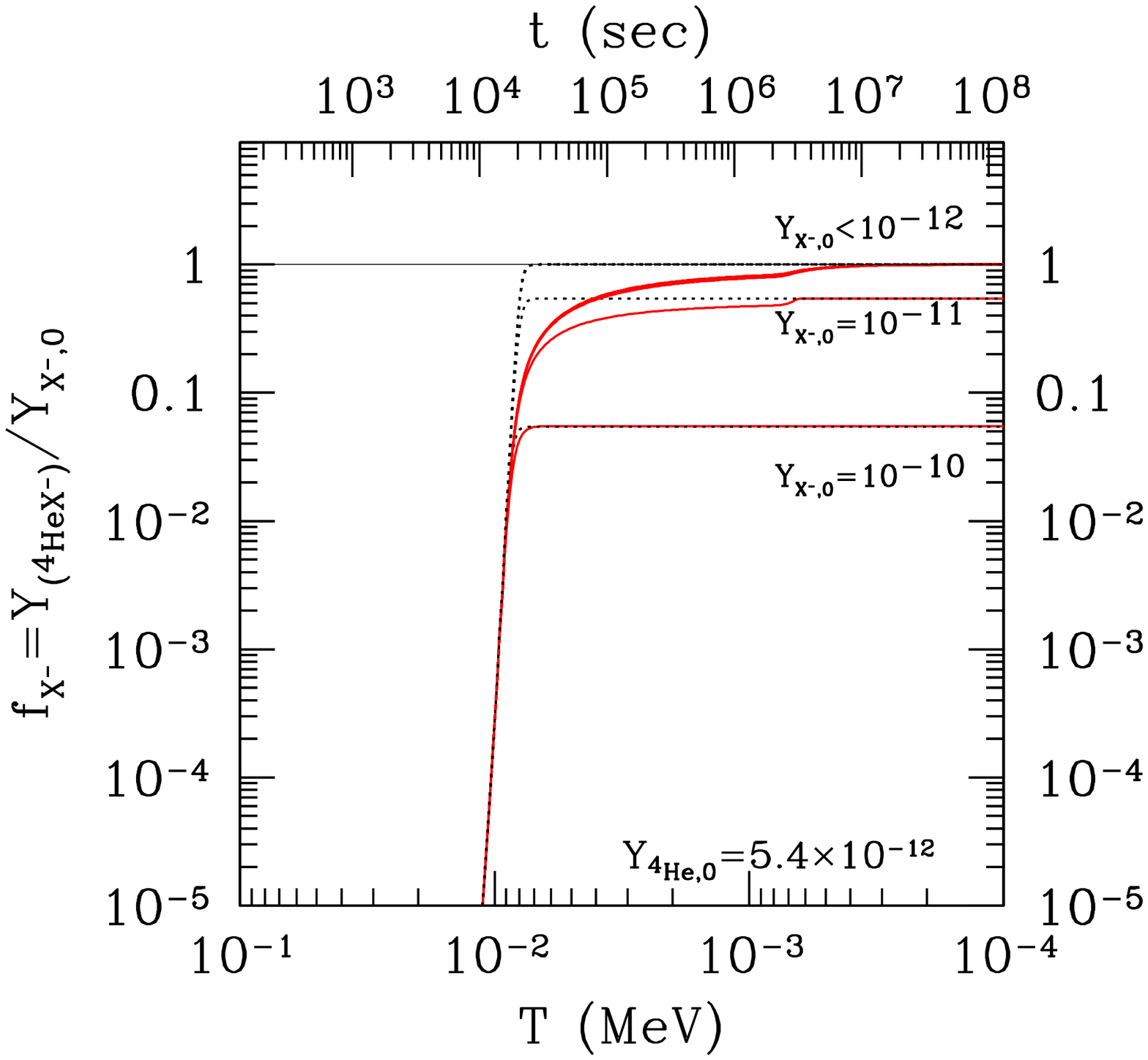}}
\resizebox{70mm}{!}{\includegraphics{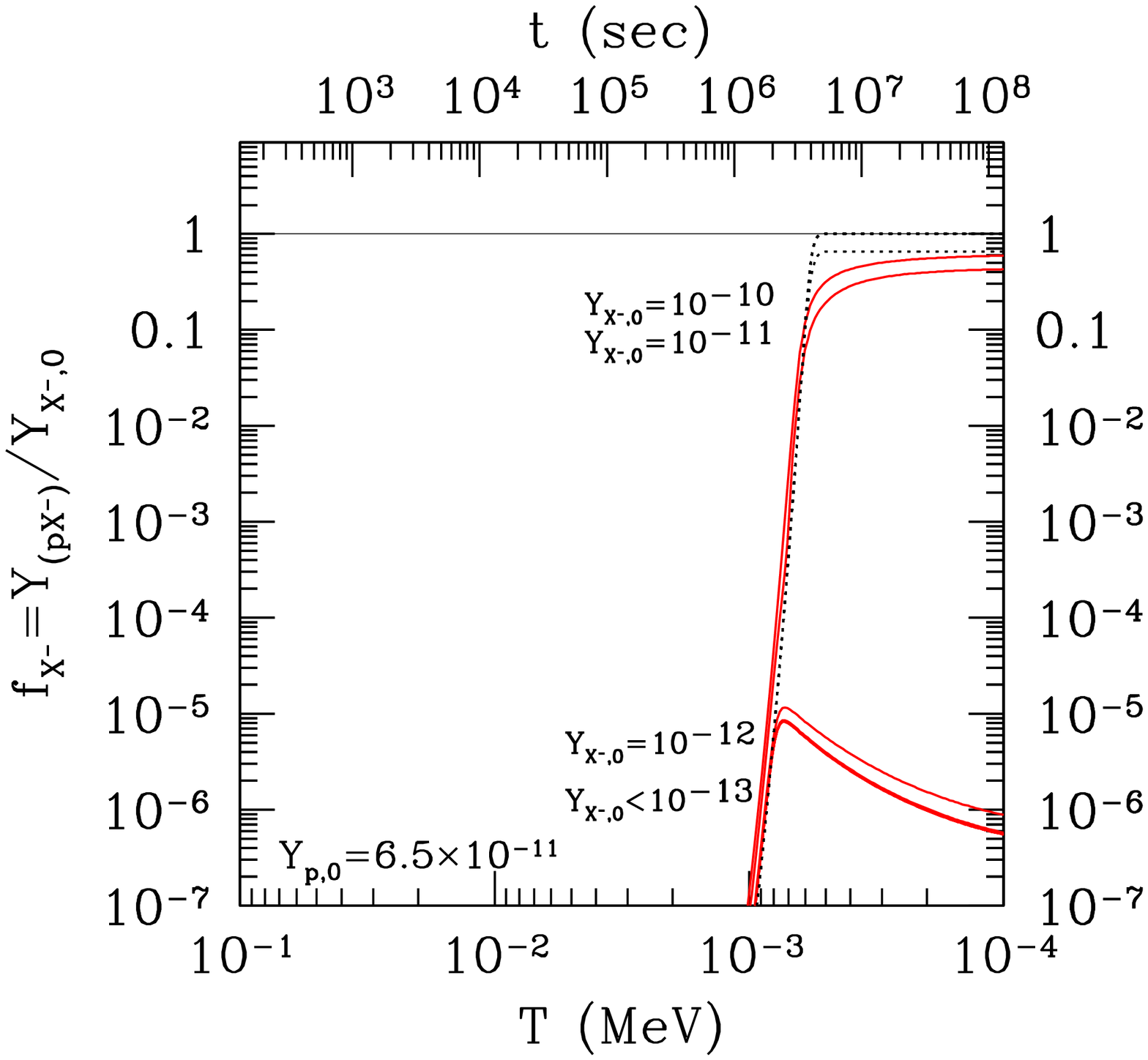}}
\caption{ Time evolution of $f_{X,N_i} \equiv Y_{(N_iX^-)}/Y_{X^-,0}$
  with $N_i=^4$He (left) and $N_i=p$ (right). The meanings of the
  labels and lines are the same as those in Fig.\ref{fig:fraction1}.}
\label{fig:fraction2}
\end{center}
\end{figure}

In Fig.~\ref{fig:fraction1}, we plot the time evolution of $f_{\rm
  bnd, N_i} \equiv Y_{(N_iX^-)}/Y_{N_i,0}$ with $N_i=^4$He (left) and
$N_i=p$ (right), where $Y_{N_i,0}$ and $Y_{X^{-},0}$ are the initial
values of $Y_{N_i}$ and $Y_{X^{-}}$, respectively. Here the initial
value of $Y_{N_i}$ means the one after $N_{i}$'s standard BBN
processes have finished and well before its bound-state formation
starts. $Y_{X^-,0}$ is the initial value of $Y_{X^-}$ well before
$X^{-}$ decays, and/or its bound-state is formed.  The dashed lines
denote solutions of Saha's equation\footnote{
  Note that Saha's equation represents the equilibrium between $X^{-}$
  and $N_{i}$ and assumes only one component of $N_{i}$. Therefore,
  there should be deviations from this solution when we consider two
  components of $N_{i} (= p$~and~$^{4}$He). However because the dashed
  lines still give us useful information to understand behaviors of
  the numerical solutions, we also plot them in
  Figs.~\ref{fig:fraction1} and \ref{fig:fraction2}
},
\begin{eqnarray}
  Y_{(N_{i}X^{-})} = 
  \left( \frac{m_{\rm ^4He} m_{X}T}{2\pi m_{(N_{i}X^{-})} }
   \right)^{-3/2} 
e^{E_{b i}/T}  s 
  ( Y_{N_{i}} - Y_{(N_{i}X^{-})} )
  ( Y_{X^{-}} - Y_{(N_{i}X^{-})} ).
  \label{SahaEq}
\end{eqnarray}
From the figure, we can see that the bound states form at around $T
\sim$ 10 keV and 1 keV for $(^{4}{\rm He}X^{-})$ and $(pX^{-})$,
respectively.

The time-evolution of the $(^{4}{\rm He}X^{-})$ formation excellently
agrees with the solution of the Saha's equation.  On the other hand,
deviations from the Saha's equation can be seen in case of $(pX^{-})$.
The behavior of the time-evolution of $(pX^{-})$ seen in
Fig.~\ref{fig:fraction1} (right) can be easily understood as
follows. If $Y_{X^{-},0} \gg Y_{^{4}{\rm He},0}$, sufficient amounts
of free $X^{-}$ exist, independently of the detail of $(^{4}{\rm
  He}X^{-})$ formation, and the abundance of $(pX^{-})$ approximately
follows the solution of Saha's equation. On the other hand, if
$Y_{X^{-},0} \ll Y_{^{4}{\rm He},0}$, the abundance of $(pX^{-})$
should deviate from Saha's equation. Furthermore, the charge-exchange
reaction becomes important when $\langle \sigma_{{\rm bnd},p} v
\rangle Y_{X^{-}} Y_{p} \sim \langle \sigma_{{\rm ex}} v \rangle
Y_{(pX^-)} Y_{\rm ^4He}$. Then the formation of $(pX^-)$ is balanced
between the production and the destruction processes, and the
abundance becomes approximately the order of $Y_{(pX^-)} \sim 0.5
\times 10^{-5} Y_{X^{-},0} (T_{c}/{\rm keV})^{-1/2}$.  Thus the
production of $(pX^{-})$ stops at around $T_{c} \sim 0.6$ keV for
$Y_{X^{-},0} < 10^{-12}$ seen in Fig.~\ref{fig:fraction1}
(right). This feature is consistent with Fig.~2.7 in
Ref.~\cite{Pradler:2009mt} and what was stated in
Ref.~\cite{Kamimura:2008fx}.  The reason why $X^-$s are included
mainly into $^4$He is that the number density of $X^-$ is much smaller
than that of electron, wihch is completely different from the case of
the standard recombination of electron.

On the other hand, in Fig.~\ref{fig:fraction2} we plot the time
evolution of $f_{X,N_i} \equiv Y_{(N_iX^-)}/Y_{X^-,0}$ with $N_i=^4$He
(left) and $N_i=p$ (right).  Similarly to Fig.\ref{fig:fraction1} the
dashed lines show the result of Saha's equation.  From
Fig.~\ref{fig:fraction2} we see that most of $X^{-}$ are captured into
the bound state with $^{4}$He if $Y_{X^-,0}$ is smaller than
$10^{-12}$. This means that ($pX^{-}$) disappears immediately after its formation
for $Y_{X^-,0} \lesssim 10^{-12}$. Then we find that almost all
$X^{-}$s are captured by $^{4}$He, and form the bound state
($^{4}$He$X^{-}$). Because the abundances of other singly-charged
nuclei such as deuterium and tritium, are much smaller than that of
proton, we can omit contributions from deuterium and tritium.  This
result tells us that the bound states of $X^{-}$ cannot be neutralized
even after ($pX^{-}$) could have been formed.

The fact that the total electric charges of the dominant
non-relativistic components cannot be shielded has a strong impact on
the structure formation. We discuss this issue in the next section.

\section{Effect on Large scale structure}\label{sec:LSS}

In this section, we discuss the effect of the charged bound state on
large scale structure.  In fact, the discussion below also applies to
charged (massive) particles themselves when they constitute a dominant
non-relativistic component. Thus in the following, we use ``CHAMPs" to
indicate both the free CHAMPs and the bound states of CHAMPs with
light elements which have a net electric charge.  When there exist
long-lived CHAMPs, they couple with photon-baryon fluid, thus matter
density fluctuations oscillate under the scales which enters the
horizon before CHAMPs decay.  Thus fluctuations of matter cannot grow,
then the matter power spectrum is suppressed on corresponding scales,
which is called ``acoustic damping" in literature.

An explicit calculation has been done on how the matter power spectrum
is suppressed in \cite{Sigurdson:2003vy,Profumo:2004qt}.  To obtain
the constraint on the decay rate rigorously, we need to compare the
matter power spectrum with observational data.  However, when CHAMPs
are dominant component of non-relativistic matter, the matter spectrum
is abruptly suppressed at the damping scale. Thus the evaluation of
the damping scale would be enough to obtain the constraint on
CHAMPs. Hence in the following, we simply make an estimate of the
acoustic damping scale as a function of the decay rate.

The scale under which matter power spectrum is suppressed, which we
denote $k_X$ in the following, can be estimated as follows.  For
fluctuations of the scale which enters the horizon before CHAMPs decay
($ t < \tau_X$ where $\tau_X$ is the lifetime of CHAMPs), they are
damped by the acoustic oscillation. On the other hand, fluctuations of
the scale which enters the horizon when $t > \tau_X$, in other words,
they can be assumed to be usual neutral dark matter, fluctuations of
such scales grow with time as in the standard case\footnote{
  Here we do not consider free-streaming of nonthermally-produced
  dark-matter particles because it depends on the kinetic energy of
  the dark matter just after its production by the $X^{-}$ decay,
  which would be highly model-dependent.  To obtain a conservative
  constraint, we neglect this effect here.
}.  Since the horizon-crossing occurs when $k = aH$ with $H$ being the
Hubble parameter, the characteristic scale $k_X$ under which the
matter power spectrum is suppressed is defined by
\begin{equation}
k_{X} \equiv  \left. aH \right|_{t=\tau_X}.
\end{equation}
We assume that the CHAMPs decay during radiation-dominated (RD)
epoch. Then $H$ is related to the cosmic time as $H = 1/2t$.  On the
other hand, during RD, $H$ can be also written as $ H^2 \simeq
\rho_{\rm rad} / 3 M_{\rm pl}^2 = H_0^2 \Omega_{\rm rad} a^{-4}.  $
Putting these together, we obtain the characteristic scale $k_X$ as
\begin{equation}
k_X = \sqrt{\frac{H_0}{2\tau_X}} \Omega_{\rm rad}^{1/4}.
\end{equation}
To make some rough estimate, we recall that $\Omega_{\rm rad} h^2
\simeq 4.15 \times10^{-5}$ and $H_0 \sim h/(3 \times 10^{17} ) ~{\rm
  s}$, we obtain
\begin{equation}
k_X \simeq 10^{4} \sqrt{\frac{\rm s}{\tau_X}} ~{\rm Mpc}^{-1}.
\end{equation}
For example, when $\tau_X \sim 1~{\rm s}$, the corresponding damping
scale is $k_X^{-1} \sim 0.1~{\rm kpc}$.  For $\tau_X \sim 10^6~{\rm
  s}$, the damping scale becomes $k_X^{-1} \sim 10^2~{\rm kpc}$, which
is the order of the galaxy scale.  Here it should be mentioned that
the damping scale down to $k_X^{-1} \sim 1~{\rm kpc}$ can be probed
with future observations of QSO-galaxy strong lens system
\cite{Hisano:2006cj}. Thus the constraints from large scale structure
would be much more stringent and such future observations give us a
lot of information on CHAMPs.

\section{Constraints on CHAMPs}\label{sec:const}

\begin{figure}[tp]
\begin{center}
\resizebox{100mm}{!}{\includegraphics{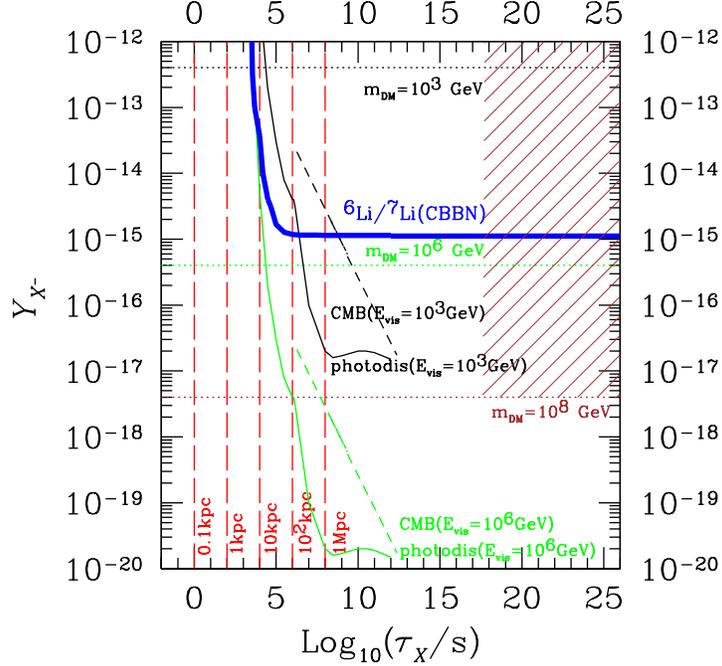}}
\caption{Constraints on the yield variable and decay rate of negative
  CHAMPs.  
The energy
density of CHAMPs is fixed to give the present dark matter density if
they are stable and then the mass is also fixed by this requirement.
The corresponding masses are given by  the horizontal
  dotted lines with the label near the lines (See also Eq.~\eqref{eq:Y_ch}). 
  The right side of vertical long-dashed line is excluded by
  the requirements that $0.1\,\mbox{kpc}$, $1\,\mbox{kpc}$,
  $10\,\mbox{kpc}$, $100\,\mbox{kpc}$, and $1\,\mbox{Mpc}$, size
  structure should not be erased, which is plotted from left to right.
  $100\,\mbox{kpc}$ corresponds to typical galaxy size
  structure. Upper region of thick solid line is
  excluded by $^{6}$ Li overproduction by the Catalyzed BBN of the
  bound-state
  effect~\cite{Pospelov:2006sc,Steffen:2006wx,Hamaguchi:2007mp,
    Kawasaki:2007xb,Bird:2007ge,Pradler-Steffen}. 
    Notice that the constraint from CBBN does not depend on the CHAMP mass. We also show the constraint from BBN and CMB
  spectrum with the mass  of CHAMPs being fixed.  
  These constraints should be interpreted with some care 
  in this figure since the mass is fixed independently of the requirement of 
  Eq.~\eqref{eq:Y_ch} for these cases.  
  Upper region of thin solid line is excluded
  by the BBN constraints from an minimal assumption of possible
  photodissociation for the visible energy of the decay, $E_{\rm vis}
  = 10^3~\mbox{GeV}$ and $10^6~\mbox{GeV}$, respectively. They are
  obtained by appropriately scaling the result
  of~\cite{Kawasaki:2000qr,Kawasaki:2004yh}. Upper region of dashed
  lines are excluded by $\mu$- and $y$-distortions of CMB spectrum for
  same visible energies.  Their labels are located near the lines.
    For the case where the lifetime is longer than the age of the
  Universe, the constraint from the deep sea water may apply, which is
  shown as the shaded region.}
\label{fig:constraint}
\end{center}
\end{figure}

Now we discuss the constraint on the abundance and the decay rate of
CHAMPs from some cosmological observations, paying particular
attention to that from large scale structure.  We show the constraint
on the $Y_{X^-}$ vs. $\tau_X$ plane, where $Y_{X^-}$ is the yield
variable of $X^{-}$ which is the number density of $X^{-}$ to the
entropy density ratio.  In the following, we assume that the energy
density of CHAMPs is fixed to give the present dark matter density if
they are stable.  (i.e., the energy density of CHAMPs before the decay
is fixed by this requirement.) Thus the yield variable and the mass
are related as
\begin{equation}
\label{eq:Y_ch}
Y_{X^-} \simeq 4 \times 10^{-12} \, \Omega_{\rm DM} h^2  \,
\left( \frac{10^3\,\mbox{GeV}}{m_{X^{-}}} \right).
\end{equation}

In Fig.~\ref{fig:constraint}, we draw the acoustic damping scales
$k_X^{-1}=1\,\mbox{Mpc}, 100\,\mbox{kpc}, 10\,\mbox{kpc},
1\,\mbox{kpc}$ and $0.1\,\mbox{kpc}$ by the vertical long-dashed
lines.  If we confirm structures larger than a scale, we can exclude
the right region of the corresponding line.  Here it should be noted
that unstable CHAMPs can erase the structure of small scales, and thus
can solve some problems regarding the discrepancies between
observations \cite{dwerf,core} and predictions of N-body simulation in
$\Lambda$CDM model~\cite{Navarro:1995iw,Moore:1999nt,Gilmore:2007fy,
  Kravtsov:2009gi,Primack:2009jr}~\footnote{
  It has been discussed that non-thermal production of warm dark
  matter can also erase the small scale structure by the long
  free-streaming length due to its relatively large velocity
  dispersion.~\cite{WDMandSSS,Hisano:2006cj}

  Furthermore, see also Refs.~\cite{Feng:2009mn,Kaplan:2009de} and
  references therein for another idea to erase small scale structures
  by introducing bound-state formation through hidden gauge
  interactions of dark matter. Note that in their models the formation
  of the bound-state means the kinetic decoupling and the end of the
  acoustic oscillation of the hidden-charged dark matter. If they
  considered hidden $^{4}$He additionally as well as standard
  cosmology, their situations might be changed.
}.

Another important constraint comes from the overproduction of $^{6}$Li
by the Catalyzed BBN through $(^{4}$He$X^{-})$ + D $\to$ $^{6}$Li +
$X^{-}$ induced by the formation of
$(^{4}$He$X^{-})$~\cite{Pospelov:2006sc}. Observational fraction of
$^{6}$Li/$^{7}$Li gives a upper bound on
$Y_{X^{-}}$~\cite{Pospelov:2006sc,Steffen:2006wx,Hamaguchi:2007mp,
  Kawasaki:2007xb,Bird:2007ge,Pradler-Steffen}.
Notice that the constraint from CBBN does not depend on the CHAMP mass.
The upper bound on $Y_{X^-}$ for $\tau_X > 10^6$~sec
is $Y_{X^-} < 2 \times 10^{-15}$. Thus, requiring the abundance of 
CHAMPs gives the present dark matter density if they are stable, 
the constraint from large scale structure becomes more relevant than that
from CBBN for $m_X \gtrsim 10^6$~GeV.

For reference, we also show other possible constraints placed on
$Y_{X^{-}}$ and $\tau_X$ from the photodissociation of BBN
\cite{Kawasaki:2000qr} and CMB $y$ and $\mu$-distortion of CMB
spectrum \cite{Hu:1993gc} with the mass of CHAMPs being fixed
independent of the relation of Eq.\,\eqref{eq:Y_ch}. Since 
the way of fixing the mass is different 
from that in other constraints, these constraints should be 
interpreted with some care in the figure.
Here we assumed that $X^{-}$ emits
electromagnetic particles and take some representative values for the
energy injected from the decay of $X^{-}$ as $E_{\rm vis} =
10^3\,\mbox{GeV}$ and $10^6\,\mbox{GeV}$.  The lines of the
constraints from the photodissociation and the CMB are obtained by
appropriately scaling the result
of~\cite{Kawasaki:2000qr,Kawasaki:2004yh} and \cite{Hu:1993gc},
respectively. In addition, here we have assumed that the branching
ratio into electromagnetic particles $(B_{\rm vis} \equiv E_{\rm
  vis}/m_{X})$ would be the order of unity. The readers can easily
obtain these bounds from photodissociation and CMB distortions by
scaling the branching ratio into electromagnetic particles
correspondingly.  When the lifetime of CHAMPs is longer than the age
of the Universe, the constraint from the sea water may apply
\cite{Smith:1979rz,Smith:1982qu,Hemmick:1989ns,Verkerk:1991jf,
  Yamagata:1993jq,Amsler:2008zzb}. In Fig.~\ref{fig:constraint}, we
fixed the energy density of CHAMPs to become the same as the present
DM density. Thus the constraint from the sea water is $Y_{X^{-}} < 4
\times 10^{-18}$, which corresponds to $m_X = 10^{8}\,\mbox{GeV}$.  As
mentioned above, the constraint from large scale structure does not
depends on the mass and the yield variable.  From the figure, we can
see that large scale structure can give the stringent constraint in
some parameter regions.

So far we have discussed the case that the dominant component of the
nonrelativistic particles in the universe could only be the negative
CHAMPs.  It would be trivial that long-lived positively-charged
particles with $\tau_{X} \gtrsim 10^{6-8}$s can be simply excluded by
the same reason from the viewpoint of the large-scale structure
without taking into account neither the bound-state formation nor
their neutralization if they become the dominant component.

\section{Summary}\label{sec:summary}

In this paper, we have investigated the evolution of the bound state
of CHAMPs with light elements and have shown that the
negatively-charged massive particle cannot be neutralized even after
its bound-state formation with protons at around 0.5 keV. This is
because the charge-exchange reaction by free $^{4}$He through
($pX^{-}$) + $^{4}$He $\to$ ($^{4}$He$X^{-}$) + $p$ is much more rapid
than the cosmic-expansion rate, and almost all $X^{-}$ will be
included into ($^{4}$He$X^{-}$) for $Y_{X^{-},0} \lesssim 10^{-12}$,
which is positively-charged.

This gives a high impact on the formation of the large-scale structure
if those charged particles are dominant non-relativistic components of
the universe like cold dark matter at the cosmic time $t \gtrsim
10^{6}\,\mbox{s}$. Then any galaxies cannot be formed by the
suppression of the density perturbation through the acoustic
oscillations.  This simply means that the lifetime of the negative
CHAMPs should be $\tau_{X} < 10^{6}$ s at longest.

As was discussed in the text, future observations of QSO-galaxy strong
lens system can probe the structure down to $k_X^{-1} \sim 1~{\rm
  kpc}$. Those future observations will reveal the nature of the
long-lived CHAMPs. 
In this letter, we have presented the result for 
the case where the abundance of CHAMPs is fixed to 
be the present-day dark matter density if they are stable.
However, it would be interesting to investigate the 
case of the energy density of CHAMPs being changed,
which will be the issue of a separate paper.

\section*{Acknowledgments}

We would like to thank Kaiki~T.~Inoue and Masayasu~Kamimura for useful
discussions.  This work is supported in part by PPARC grant
PP/D000394/1 (K.K.), and Grant-in-Aid for Scientific research from the
Ministry of Education, Science, Sports, and Culture, Japan,
No. 19740145 (T.T.) and No. 18071001 (K.K.).


\end{document}